\documentstyle[preprint,aps,epsfig]{revtex}
\def\ga{\mathrel{\raise.3ex\hbox{$>$\kern-.75em\lower1ex\hbox{$\sim$}}}}
\def\la{\mathrel{\raise.3ex\hbox{$<$\kern-.75em\lower1ex\hbox{$\sim$}}}}
\begin{document}

\newcommand{\nn}{\noindent}
\renewcommand{\thefootnote}{\fnsymbol{footnote}}

\begin{flushright}
KIAS Preprint P01025 \\
UFR-HEP/0105\\
July 2001 \\
\end{flushright}
\vspace{0.5cm}
\begin{center}
{\Large{ {\bf  Probing scalar--pseudoscalar mixing 
in the CP violating MSSM at high--energy $e^+e^-$ colliders}}}

\vspace{.4cm}

{\large A.G. Akeroyd}$^{\mbox{1}}$\footnote{E--mail: akeroyd@kias.re.kr},
{\large A. Arhrib}$^{\mbox{2}}$
\footnote{On leave of 
absence from Department of Mathematics, FSTT, P.O.B416 Tangier, Morocco.
 E--mail: arhrib@phys.ntu.edu.tw}

\vspace{.4cm}
{\sl
1: Korea Institute for Advanced Study, 207--43 Cheongryangri--dong,\\ 
Dongdaemun--gu, Seoul 130-012, Korea

\vspace{.3cm}
2: Physics Department, National Taiwan University\\
10764 Taipei, Taiwan, R.O.C} 
   
\end{center}

\setcounter{footnote}{4}
\vspace{0.2cm}

\begin{abstract}
\nn 
We study the production processes $e^+e^-\to 
H^0_iZ$, $H^0_iH^0_j$ and $H^0_i\nu_e\overline \nu_e$
in the context of the CP violating MSSM. 
In a given channel we show that the cross--section for all $i$
($=1,2,3$) can be above 0.1 fb provided $M_{H_{2,3}}\la 300$ GeV. 
This should be detectable at a Next Linear
Collider and would provide evidence for scalar--pseudoscalar mixing.

\end{abstract}
\vfill

\newpage

\pagestyle{plain}
\renewcommand{\thefootnote}{\arabic{footnote} }
\setcounter{footnote}{0}

\section*{1.~Introduction}
The search for the scalar particles referred to as
``Higgs bosons'' \cite{Higgs} is one of the major goals of present and 
future colliders. Such particles break the electroweak symmetry
and are responsible for the masses of the fermions and bosons.
The Standard Model (SM) \cite{Wein} predicts one neutral scalar ($\phi^0$)
while extensions of the SM often predict several scalars, both
neutral and charged \cite{gun}.
Since 1989 the $e^+e^-$ collider LEP has searched for 
$\phi^0$. In the final run at energies around $\sqrt s=208$ GeV two of the 
four experiments found excesses in the search for a Higgs boson in 
association with a Z boson \cite{evid}. 
Although it was agreed 
that a further run with about 200 pb$^{-1}$ per experiment at a 
centre--of--mass energy of 208.2 GeV would enable the four experiments to
establish a 5$\sigma$ discovery (assuming the signal is genuine)
\cite{www}, the extended run was not approved and LEP was consequently 
shut down. The search for Higgs bosons will continue with Run II at 
the Tevatron \cite{tevatronH} which has a chance of confirming the existence 
of the Higgs boson in the mass range hinted at by LEP ($\approx 115$ 
GeV). This region is fairly problematic for the LHC \cite{LHC}
and would require several years searching in the channel $h\to \gamma\gamma$
to confirm such a light Higgs. A higher--energy $e^+e^-$
collider (NLC) operating at energies $\sqrt s \ge 500$ GeV would
be an ideal place to perform precision measurements of a Higgs boson 
\cite{NLC},\cite{eecolliders}.

Supersymmetric (SUSY) theories, in particular the Minimal Supersymmetric
Standard Model (MSSM) \cite{MSSM}, are currently considered as the 
most theoretically well motivated extensions of the SM. Such models 
predict a rich Higgs
phenomenology. The MSSM contains 2 neutral CP--even scalars ($h^0$ and $H^0$),
a CP--odd neutral scalar $A^0$, and a pair of charged scalars 
($H^+,H^-$). High--energy $e^+e^-$ colliders offer a very clean environment
in which to search for Higgs bosons, and the simplest way to produce 
a CP--even scalar is in the Higgsstrahlung process 
$e^+e^-\to Z^*\to H^0Z$ and the $W$ boson fusion process
$e^+e^- \to H^0\nu_e\overline \nu_e$ \cite{HZ,2HZ}.
The CP--odd $A^0$ possesses no tree--level coupling $A^0VV$ (where
$V=Z,W^\pm$) and the other tree--level diagrams contributing to
$e^+e^-\to A^0Z$ are proportional to the electron mass and 
consequently negligible\footnote{Note that at a muon collider 
\cite{aad}, the tree--level diagrams for $e^+e^-\to A^0Z$ 
cannot be discarded anymore and higher order diagrams would induce
corrections to the tree-level rate.}. 
Therefore the dominant contribution to 
$e^+e^-\to A^0Z$ is from higher order diagrams.
In previous works we calculated the 1--loop induced rate in the 
context of the CP--conserving THDM \cite{AAC} and the CP--conserving MSSM 
\cite{MSSMAAC}, finding maximum values of order 0.01 fb and 0.1 fb 
respectively if $\tan\beta\ge 2$.
The Higgsstrahlung, vector boson fusion, and
Higgs pair production mechanisms 
($e^+e^- \to A^0h^0/H^0$ and $e^+e^-\to h^0H^0$), 
have all been extensively studied (including radiative corrections
\cite{hollik,chankowski,djouadi}) at NLC energies in the context of the MSSM.

In recent years the phenomenology of the MSSM with complex SUSY parameters
has received growing attention \cite{MSSMCP},
\cite{MSSMCP2}, \cite{plb495},\cite{MSSMCP1}. 
Such phases may allow baryogenesis \cite{Baryogen}, and do
not necessarily violate the stringent bound from the non--observation
of Electric Dipole Moments (EDMs) \cite{nath}, \cite{EDM}.
The presence of SUSY phases induces mixing between the CP--even and 
CP--odd scalars, resulting in the mass eigenstates $H_1^0$,
$H_2^0$ and $H^0_3$ which are mixed states of CP. 
This mixing affects their phenomenology at present
and future colliders, both in production mechanisms and decay partial 
widths \cite{BR}. In this paper we will study the processes
$e^+e^-\to Z^*\to H^0_iZ$ and $e^+e^-\to H^0_i\nu_e\overline \nu_e$
in the context of a NLC. Both mechanisms are mediated by the tree--level
couplings $H^0_iVV$, but their cross--sections have different 
phase space and $\sqrt s$ dependence. In the CP conserving case one of
these couplings would be zero, corresponding to the absence of 
the coupling $A^0VV$. We will also study the production of 
neutral Higgs pairs, $e^+e^- \to H^0_iH^0_j$ ($i\neq j$). In the CP 
conserving MSSM, only the vertices $Zh^0A^0$ and $ZH^0A^0$ exist at 
tree--level while in the CP violating scenario all three couplings 
$ZH^0_1H^0_2$, $ZH^0_1H^0_3$ and $ZH^0_2H^0_3$ are generated at tree--level.
Therefore an observable signal for all $i(=1,2,3)$
in a given mechanism ($e^+e^-\to ZH_i^0$, $e^+e^-\to H_i^0 H_j^0$ or  
$e^+e^-\to H^0_i\nu_e\overline 
\nu_e$), would be a way of probing CP violation in
the Higgs sector. Such an approach was used in the context of
the THDM in \cite{Pomarol}, with related analyses in \cite{GunionKal}. 
We will calculate the tree--level 
rates of the above mechanisms in the context of the CP violating 
MSSM, showing that in the most favourable scenarios this 
way of probing scalar--pseudoscalar mixing can be effective
if $M_{H_{2,3}}\la 300$ GeV. 

Our work is organized as follows. In section 2 we outline our
approach for evaluating the cross--sections for 
the above production mechanisms in the CP violating MSSM. In section 3 we 
present our numerical results and section 4 contains our conclusions.

\renewcommand{\theequation}{4.\arabic{equation}}
\setcounter{equation}{0}
\section*{2. Scalar-Pseudoscalar mixing in the MSSM}
The tree--level Higgs potential of the MSSM conserves CP, which ensures that
the three neutral Higgs eigenstates can be divided into the CP--even
$h^0$ and $H^0$ and CP--odd $A^0$. Recent studies 
\cite{MSSMCP,MSSMCP2,MSSMCP1} 
have shown that the 1--loop effective potential may violate CP resulting 
in three Higgs mass eigenstates which cannot be assigned a definite 
CP quantum number, denoted by $H^0_1$,$H^0_2$ and $H^0_3$ (in ascending
order of mass).
In the above studies it is shown that the  CP violation is generated 
by complex phases which reside in
the $\mu$ term and the soft SUSY breaking parameters 
$A_t$ and $A_b$. These phases generate terms ${\cal M}^2_{SP}$ in
the $3\times 3$ neutral Higgs mass squared matrix ${\cal M}^2_{ij}$ 
which mix the CP--odd and CP--even scalar fields. These may be given
approximately by
\cite{MSSMCP}
\begin{eqnarray}
{\cal M}^2_{SP}\approx {\cal O}\left({m^4_t|\mu||A_t|
\over v^232\pi^2M^2_{SUSY}}\right)\sin\phi_{CP}
\times \left[6,{|A_t|^2\over M^2_{SUSY}},{|\mu|^2\over \tan\beta M^2_{SUSY}},
{\sin2\phi_{CP}|A_t||\mu|\over \sin\phi_{CP}M^2_{SUSY}}\right] 
\end{eqnarray}
where $\phi_{CP}={\rm arg}(A_t\mu)$, and we have only displayed the
contributions from the top squarks, $\tilde t_{1,2}$, 
which are dominant for small $\tan\beta$. 
Sizeable scalar--pseudoscalar mixing
is possible for large $|\mu|$,$|A_t| > M_{SUSY}$.
In  \cite{MSSMCP1} the mass matrix 
${\cal M}^2_{ij}$ is evaluated 
to one--loop order using the effective potential techniques and includes 
large two--loop non--logarithmic corrections induced by one-loop threshold 
effects on the top and bottom quark Yukawa coupling. The public code
which we will employ in our numerical analysis  can be found 
in \cite{cph}. 
If the SUSY phases are set to zero, the mass eigenstates become 
definite eigenstates of CP. A phenomenological consequence of
the scalar--pseudoscalar mixing is that
all the eigenstates $H^0_i$ possess a tree--level 
coupling $VVH^0_i$ with respective strength $C_i$ (normalized to SM strength).
These couplings can be easily obtained from the 
covariant derivative of the Higgs fields in which the non--physical Higgs 
fields are expressed in terms of the physical mass eigenstates.
Following the convention of \cite{MSSMCP1}, the $C_i$ are given by:
\begin{eqnarray}
C_i= O_{1i}\cos\beta+ O_{2i}\sin\beta  \label{coupl}
\end{eqnarray}
Here $O_{ij}$ is the orthogonal matrix which diagonalizes 
${\cal M}^2_{ij}$. One can easily show that
\begin{equation}
C_1^2+C_2^2+C^2_3=1\label{sum}
\end{equation}
Note that this sum rule applies to the tree--level vertices,
although it holds to a very good approximation if higher order 
corrections to the vertices are included. We will present
results using the tree--level values for $C_i$ which we
will generate by use of the program cph.f \cite{cph}. 


In the CP conserving MSSM 
scalar--pseudoscalar mixing is absent. In this case
$C_1=\sin(\beta-\alpha)$, while one of $C_2$,$C_3$
is identified as $\cos(\beta-\alpha)$, and the other is identically
zero. $C_i\equiv 0$ corresponds to the couplings $A^0ZZ$ 
and $A^0WW$, 
which will take on a non--zero value at 1-loop. 
Hence to lowest order in the MSSM only $H^0_1$ and {\it one} of $H^0_2$,
$H^0_3$ can be produced in the Higgsstrahlung 
and $WW$ fusion mechanisms, 
$e^+e^-\to Z^*\to H^0_iZ$ and $e^+e^- \to H^0_i\nu_e\overline \nu_e$. 
In the presence of SUSY phases 
all $H^0_i$ may be produced at tree--level via $e^+e^-\to Z^*\to H^0_iZ$,
and an observable signal for all three $H^0_i$ would be evidence for CP 
violation in the Higgs sector.
We stress here that the smallest of 
$\sigma(e^+e^-\to Z^*\to H^0_iZ)$ should exceed the 
maximum rate for $e^+e^-\to A^0Z$ in the 
context of the CP conserving MSSM \cite{MSSMAAC}, since the latter
would constitute a ``background'' to any interpretation of 
scalar--pseudoscalar mixing. Note that the process 
$e^+e^- \to H^0_i\nu_e\overline \nu_e$ proceeds via
the same couplings $C_i$, but possesses
a different phase space and $\sqrt s$ dependence. 
This mechanism is competitive with the
Higgsstrahlung process, and becomes the dominant one as
$\sqrt s$ increases. We will also consider the mechanism
$e^+e^-\to H^0_j H^0_k$, which proceeds via the coupling
$C_{jk}$, where $C_i=C_{jk}$ for $i\ne j\ne k$.

In the MSSM (with or without SUSY phases),  
the properties of the lightest eigenstate $H^0_1$ become
very similar to that of the SM Higgs boson in
the decoupling region of $M_{H^\pm}\ge 200$ GeV.  
In this region, $C_1$ is very close to 1, and so the sum $C_2^2 + C^2_3$ 
is constrained to be small (Eq. (\ref{sum})). Therefore we expect that
$M_{H^\pm}\le 200$ GeV will allow larger values for the sum 
$C_2^2 + C^2_3$, and thus observable rates for both $H^0_2$ and
$H^0_3$ in the above mechanisms. 
Following the approach of \cite{GunionKal} we will take the threshold of
observability as $\sigma_{obs}=0.1$ fb. This would give 50 raw events 
(before cuts) for the assumed luminosities of 500 $fb^{-1}$. 
Distinct signals for all three $H^0_i$ in a given channel 
would be evidence for scalar--pseudoscalar mixing.
Note that \cite{MSSMAAC} found maximum values of 
$\sigma(e^+e^-\to A^0Z)=0.1$ fb in the context of the CP conserving MSSM.

A caveat here is that extended Higgs sectors with more than two doublets
or extra Higgs singlets (e.g. the NMSSM) would also predict
multiple signals in these mechanisms \cite{GunionKal}. We will show 
that the CP violating MSSM can only produce multiple signals 
below a certain mass for
$M_{H_2}$ and $M_{H_3}$, and thus any such signal for a larger Higgs
mass would be evidence against the CP violating MSSM. Therefore the
measurement of the Higgs mass may act as a discriminator among the models.

\renewcommand{\theequation}{3.\arabic{equation}}
\setcounter{equation}{0}
\section*{3. Numerical results and discussion}
We now present our numerical results which we will generate with 
the fortran program cph.f \cite{cph}. We note that this program
does not include $\chi^+$--$W$--$H^\pm$ contributions to the 1--loop
neutral Higgs mass matrix, which have been shown to be sizeable in 
some regions of parameter space \cite{nathibrahim}. 
The analytic expressions 
for the various cross--sections are given in the literature \cite{NLC}.
For the $WW$ fusion process we will use the exact 
expression given in \cite{kilian}.

Graphs showing the numerical values of the couplings 
$C_i$ and $C_{ij}$ have appeared in Refs \cite{MSSMCP,MSSMCP1}. 
However, these papers were more concerned with light $M_{H_2}$
and $M_{H_3}$ of interest at LEP2 and the Tevatron. 
Detection at these colliders would require quite sizeable values 
for $C^2_2$ and $C^2_3$. We are concerned with
a NLC collider which has the ability to probe 
$\sigma(e^+e^-\to H^0_iZ,H^0_iH^0_j,H^0_i\nu_e\overline \nu)\ge 0.1$ fb, 
and so we are also interested
in smaller values for $C^2_2$,$C^2_3$ and larger $M_{H_i}$. 
An earlier discussion of the potential of
a NLC to probe scalar--pseudoscalar 
mixing can be found in \cite{demir}. However
this study only addressed the pair production process 
$e^+e^-\to H^0_i H^0_j$ and numerical results
were only presented for the couplings $ZH^0_iH^0_j$.
We shall be presenting results for the production cross--sections
of all the above mechanisms, in contrast to \cite{MSSMCP,plb495,MSSMCP1} 
which were more concerned with the numerical values of $C_i$ and $C_{ij}$
and LEP2 phenomenology. Since the $WW$ fusion process is considerably
more important at NLC energies than at LEP2 energies, our analysis
is complementary to that in \cite{plb495} and extends that
of \cite{demir}. For alternative ways of probing CP violating SUSY
phases at $e^+e^-$ colliders, see \cite{Barger:2001nu}.

As noted in the introduction, the cross--sections for
the processes $e^+e^- \to Zh^0/H^0$, 
$e^+e^- \to Ah^0/H^0$, $e^+e^- \to h^0 H^0$, $e^+e^- \to ZA^0$ and 
$e^+e^- \to \nu_e\overline \nu_e h^0/H^0$ in the CP conserving MSSM
are accurately known \cite{AAC,hollik,chankowski,djouadi}.
Deviations from these rates would be evidence 
for scalar--pseudoscalar mixing.

The presence of large SUSY phases can give contributions to 
the EDM which exceed the experimental upper bound. To avoid
conflict with experiment one may
assume that the masses of first two generation of squarks 
are well above the TeV scale while the third generation may be 
relatively light ($\le 1$ TeV) \cite{nath}. A recent paper \cite{mix} 
suggests that sizeable scalar--pseudoscalar mixing would prefer the 
cancellation mechanism \cite{EDM} over the above mechanism 
as the solution to keep the value 
of EDM within the experimental limits. Another option
is to adopt a non--universal scenario for the tri--linear couplings
$A_f$ \cite{trilin}. In particular, one may require 
${\rm arg}(\mu)\le 10^{-2}$ and $A_f=(0,0,1)A$, with $A_t$,$A_b$ and 
$A_{\tau}$
taking maximal phases. Such a scenario comfortably satisfies
the EDM constraints. Since the scalar--pseudoscalar mixing 
$\sim \phi_{CP}={\rm arg}(A_t\mu)$, it is sufficient to have maximal phase
in $A_t$ to maximize $\phi_{CP}$. However, two--loop Barr-Zee type
diagrams \cite{2-loop} can violate the EDM constraints for large
$\tan\beta$ ($\ge 30$). Therefore we will restrict ourselves to 
low to intermediate values of $\tan\beta$. 

In our numerical analysis we will choose the CP violating
benchmark scenario (CPX) which was introduced in \cite{plb495}
and maximizes the CP violating effects.
The CPX scenario is as follows:
\begin{eqnarray}
& & \widetilde{M}_Q=\widetilde{M}_t=\widetilde{M}_b=M_{SUSY}=0.5\to 1 
{\mbox{TeV}} \ , \ 
\mu = 4 M_{SUSY} \nonumber \\
& & |A_t|=|A_b|= 2 M_{SUSY} \ , \ |m_{\widetilde{g}}|= 1 
{\mbox TeV} \ {\mbox{and}}
\ |m_{\widetilde{B}}|= |m_{\widetilde{W}}|= 0.3 {\mbox TeV} 
\end{eqnarray}
Note that $\mu$ will be taken real while we allow a CP phase 
in the soft tri-linear parameters $A_t$ and $A_b$ and  
in $m_{\widetilde{g}}$. The CP phases of $A_t$ and $A_b$
are chosen to be equal and may be maximal. In addition,
we choose the charged Higgs mass and $\tan\beta$ as free parameters.

Our strategy to probe the scalar--pseudoscalar mixing 
requires the identification of the Higgs signals as distinct 
resonances. The inclusion of the phases in $A_t$ and $A_b$ breaks 
the near degeneracy among $M_{H_2}$ and $M_{H_3}$ \cite{MSSMCP1}, 
and gives sufficient splittings to allow identification of 
separate resonances for $H^0_2$ and $H^0_3$. These splittings may
be $>10$ GeV, which is sufficiently large
for a NLC \cite{NLC} to resolve the separate peaks. 
This will lead to 
three different peaks in the Higgsstrahlung and $WW$ fusion
processes and motivates us to 
present the individual cross--sections for $e^+e^- \to Z H^0_{i}$ and 
$e^+e^- \to \nu\overline{\nu}_e H^0_{i}$ for $i=1,2,3$.
This is in contrast to \cite{ham} where the study was devoted to 
LEPII energies and the cross--sections were summed over the 
three Higgs states. It has been shown in \cite{MSSMCP,plb495,MSSMCP1}
that the inclusion of SUSY phases may drastically change the size of 
the couplings $ZZH^0_1$ and $ZH^0_1H^0_2$ for low and 
intermediate $\tan\beta$. 
In such cases the bound on the light Higgs boson obtained at LEPII
may be weakened to $\la 60 $ GeV for large CP violation in the MSSM
Higgs sector. We study the potential of a NLC to discover such a weakly 
coupled Higgs.

In Fig. 1 the left (right) plots depict regions of 
$\sigma(e^+e^- \to ZH^0_{i}$) in the plane ($M_{H_i}$,{\rm arg}($A_t$))
for $\sqrt s=500$ GeV, $\tan\beta=6(15)$, and $M_{SUSY}=1000(500)$ GeV.
In all plots the charged Higgs mass has been varied in increments 
from 140 $\to$ 400 GeV, which determines the values of $M_{H_i}$.
Comparing the left and right plots it is clear that 
lower $\tan\beta$ provides larger cross--sections for 
$H^0_2$ and $H^0_3$ over a wider region of the plane, corresponding to
the fact that the scalar--pseudoscalar mixing is 
enhanced. For $H^0_1$ discovery is 
possible over most of the ($M_{H_1}$,{\rm arg}($A_t$)) 
plane, with small unobservable regions where 
$\sigma(e^+e^- \to Z H^0_{1})< 0.1$ fb which occur
for ${\rm arg}(A_t)\approx 1.5(2)$ for $\tan\beta=6(15)$ and 
$M_{H_1}\la 105(115)$ GeV.
The smallness of $\sigma(e^+e^- \to Z H^0_{1})$ is 
due to the suppression of $C_1$. This suppression 
arises when $O_{21}$ changes sign, which induces destructive
interference in Eq. (\ref{coupl}). 
Note that $C_1$ is dominated by $O_{21}$ and so $C_1$ also flips sign 
for $\tan\beta=6$(15) and ${\rm arg}(A_t)\approx 1.5$ 
($\approx 2$).
$C_2$ is positive in both cases $\tan\beta=6$ and $\tan\beta=15$,
and is maximized for $M_{H_2}\la 150$ GeV; it is minimized for  
${\rm arg}(A_t) \approx 1.5(\ga 2)$ and 
$M_{H_2}\ga 150$ GeV for $\tan\beta=6(15)$. For $H^0_2$ and $H^0_3$, 
both $\sigma(e^+e^- \to Z H^0_{2,3}$) can be observable over a wide
region of the plane, even up to relatively large mass values 
e.g. for $\tan\beta=6$ and ${\rm arg}(A_t)=1$,
$\sigma(e^+e^- \to Z H^0_{2,3})\ge 0.1$ fb for $M_{H_2}\le 250$ GeV
and $M_{H_3}\le 270$ GeV. Note that the scalar--pseudoscalar 
composition of $H^0_2$ and $H^0_3$ can change with increasing $M_{H_i}$
e.g. for the $\tan\beta=6$ plot with low ${\rm arg}(A_t$) (i.e. small
scalar--pseudoscalar mixing) one can see
that $H^0_3$ is dominantly scalar for low masses, and has a
much larger cross--section than for that for $H^0_2$. 
As $M_{H_{2,3}}$ increases, $H_2^0$ has the larger scalar
component and may be produced with an observable rate 
for $M_{H_2}\le 300$ GeV.
The coverage for $e^+e^-\to H^0_i\nu_e\overline \nu_e$
at the same $\sqrt s$ is comparable to that in Fig.1.

Fig. 2 shows $\sigma(e^+e^- \to Z H^0_{i}$) for $\sqrt s=800$ GeV,
$\tan\beta=6$, and $M_{SUSY}=1000$ GeV. Ones sees that the coverage is
inferior to that in Fig.1 since $\sigma(e^+e^- \to Z H^0_{i}$) 
is reduced for larger $\sqrt s$.

Fig. 3 is analogous to Fig.1 and shows
$\sigma(e^+e^-\to H^0_i\nu_e\overline \nu_e)$ for $\sqrt s=800$ GeV.
Here we find improved coverage compared to that in Figs. 1 and 2, since
the cross--section for this process is enhanced with increasing $\sqrt s$.
For the $\tan\beta=6$ plot with ${\rm arg}(A_t)=0.5$, both $\sigma(e^+e^-\to 
H^0_{2,3}\nu_e\overline \nu_e)\ge 0.1$ fb for $M_{H_{2,3}}\le 300$ GeV.
The window of unobservability for $H_1^0$ has essentially been closed.

In Fig. 4 we show $\sigma(e^+e^- \to H^0_iH^0_j$) in the plane
$(M_{H_i},{\rm arg}(A_t))$ for $\tan\beta=6$, $M_{SUSY}=1000$ GeV, 
and $\sqrt s=500$ GeV. This mechanism offers comparable cross--sections to
those for the Higgsstrahlung and $WW$ fusion processes, and 
consequently is also effective at probing scalar--pseudoscalar mixing.
The best coverage is obtained at ${\rm arg}(A_t)\approx 0.5$.
Using the fact that $C_i^2=C_{jk}^2$ for $i\neq j \neq k$,
the behaviour of pair production $e^+e^- \to H^0_i H^0_j$
can be roughly understood from the rate of the Higgsstrahlung process 
$e^+e^- \to Z H^0_k$. As can be seen from the plots,
there are some similarities between $e^+e^-\to H^0_i H^0_j$ and 
$e^+e^-\to Z H^0_k$, for $i\neq j \neq k$.
Note that $\sigma(e^+e^- \to H^0_1H^0_{2,3}$) have large cross--sections
($\ge 5$ fb) in the region where the Higgsstrahlung and $WW$ fusion processes
have very suppressed rates. This situation corresponds to
strong scalar--pseudoscalar mixing, since in the absence of SUSY phases
one of $\sigma(e^+e^- \to H^0_1H^0_{2,3}$) would be zero.

\renewcommand{\theequation}{4.\arabic{equation}}
\setcounter{equation}{0}
\section*{4. Conclusions}
We have studied the production processes $e^+e^-\to 
H^0_iZ$, $H^0_iH^0_j$ and $H^0_i\nu_e\overline \nu_e$
in the context of the CP violating MSSM. 
We showed that in a given
channel the cross--section for all $H^0_i$ ($i=1,2,3$) can be observable
at a Next Linear Collider and would provide evidence for 
scalar--pseudoscalar mixing. At $\sqrt s=500$ GeV the coverage of 
$e^+e^-\to H^0_iZ$ and $H^0_i\nu_e\overline \nu_e$ are comparable, with
observable cross--sections for $M_{H_2}\le 250$ GeV and 
$M_{H_3}\le 270$ GeV for the most favourable choice of ${\rm arg}(A_t)$.
At $\sqrt s=800$ GeV, the process  $e^+e^-\to H^0_i\nu_e\overline \nu_e$ 
offers superior coverage,
with a reach up to $M_{H_{2,3}}\le 300$ GeV in the most favourable cases.
The scalar--pseudoscalar mixing causes a mass splitting 
between $H^0_2$ and $H^0_3$ which should be sufficient for separate peaks
to be resolved at a NLC.
The problematic region of a light $H^0_1$ with a very suppressed coupling
to vector bosons ($VVH^0_1$) has a window of unobservability at 
$\sqrt s=500$ GeV
and ${\rm arg}(A_t)\approx \pi/2$. This is almost closed at $\sqrt s=800$ GeV
in the $H^0_i\nu_e\overline \nu_e$ channel. The mechanism
$e^+e^-\to H^0_iH^0_j$ is competitive with the
above mechanisms for probing scalar--pseudoscalar mixing at
$\sqrt s=500$ GeV, and can comfortably detect $H^0_1$ in the 
region of suppressed coupling $VVH^0_1$.

\section*{Acknowledgement}
A.~Arhrib is supported by National Science
Council under the grant NSC 89-2112-M-002-063.

\renewcommand{\theequation}{B.\arabic{equation}}
\setcounter{equation}{0}

\newpage

\begin{figure}[t!]
\centerline{{
\epsfxsize3.8 in 
\epsffile{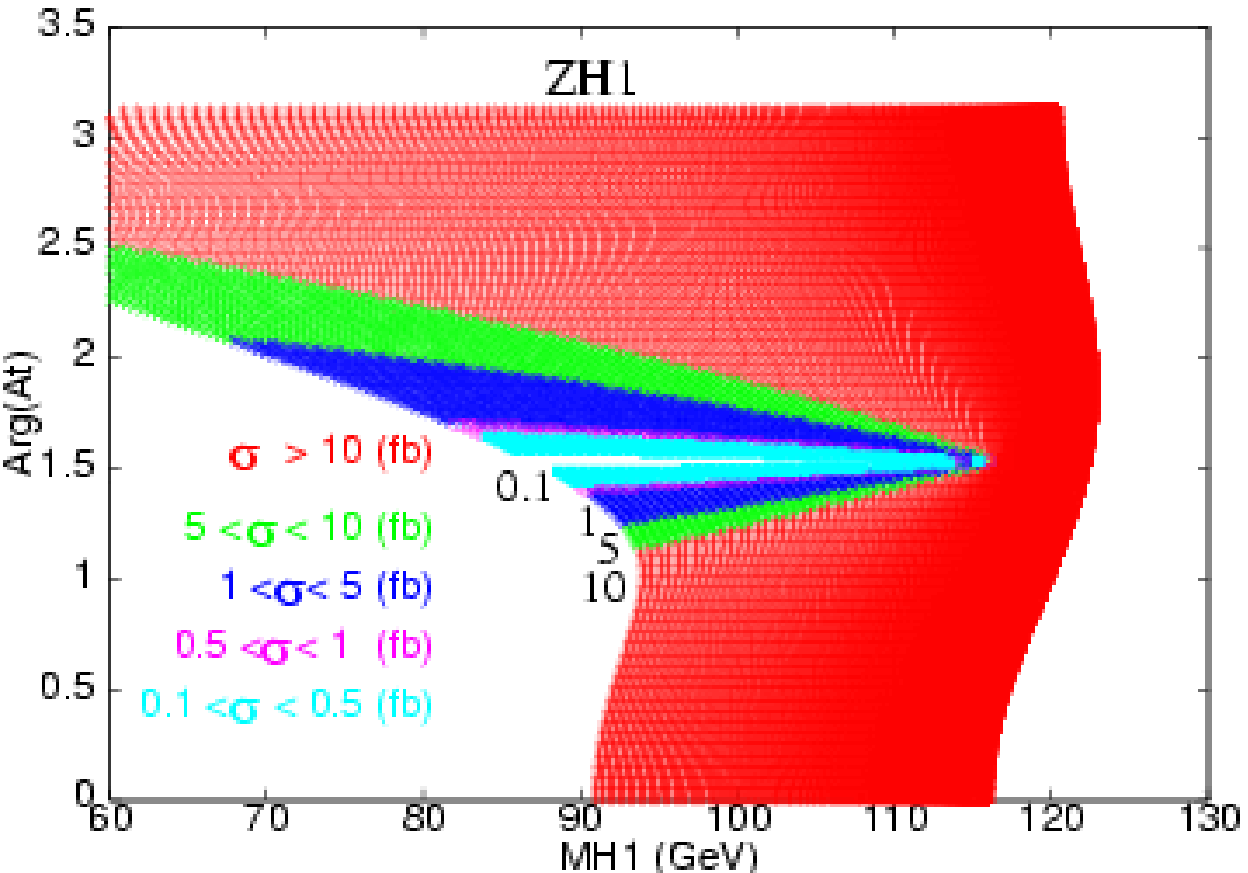}}
\hskip0.4cm
            {\epsfxsize3.8 in \epsffile{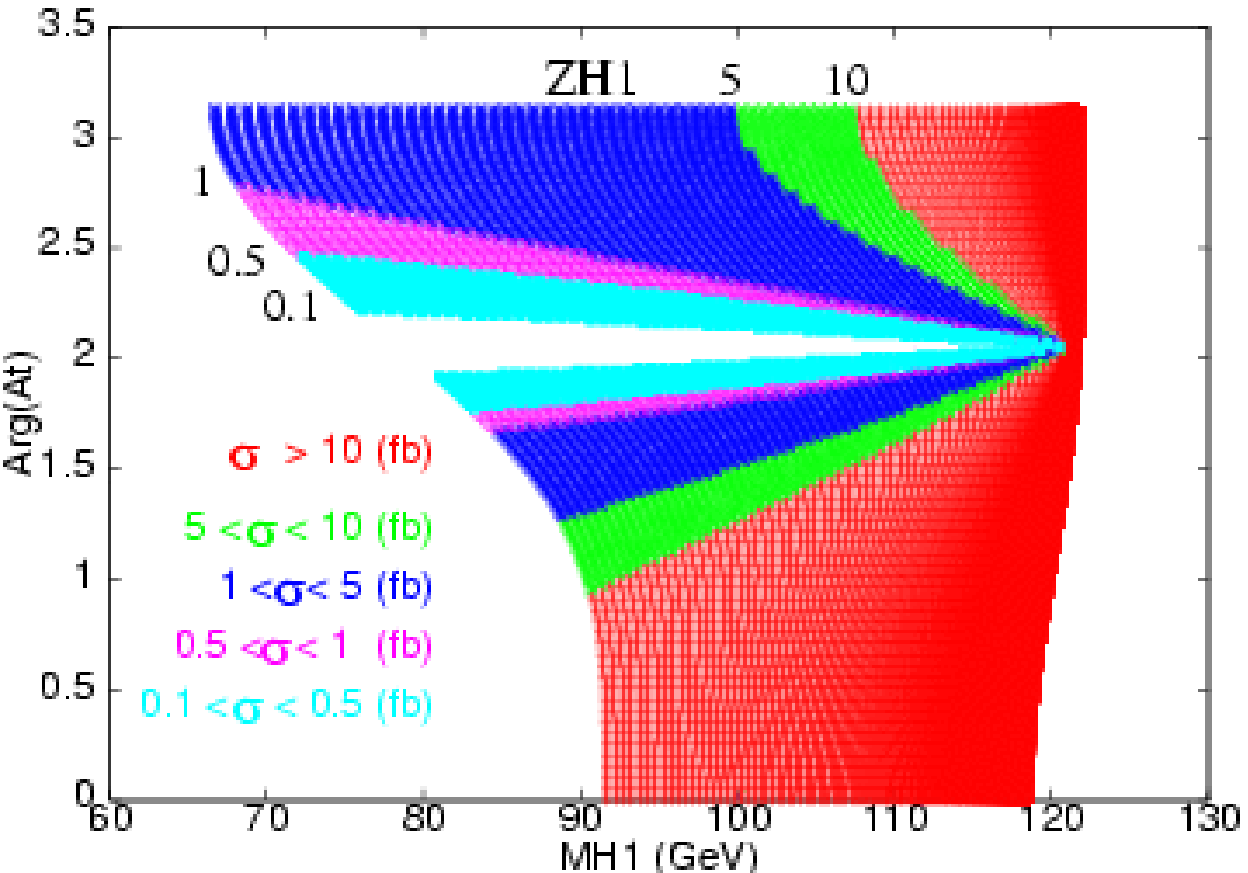}}}
\smallskip\smallskip
\centerline{{
\epsfxsize3.8 in 
\epsffile{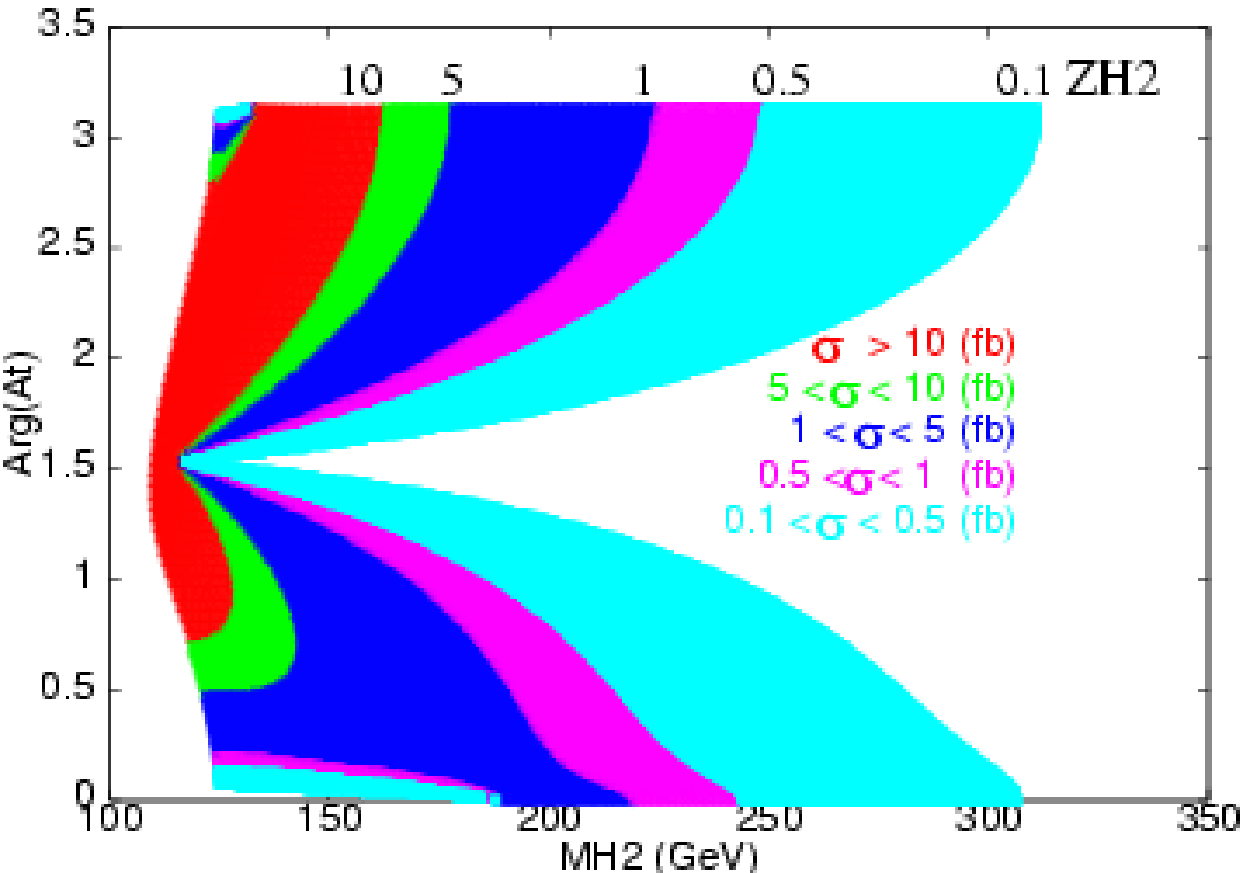}}
\hskip0.4cm
            {\epsfxsize3.8 in \epsffile{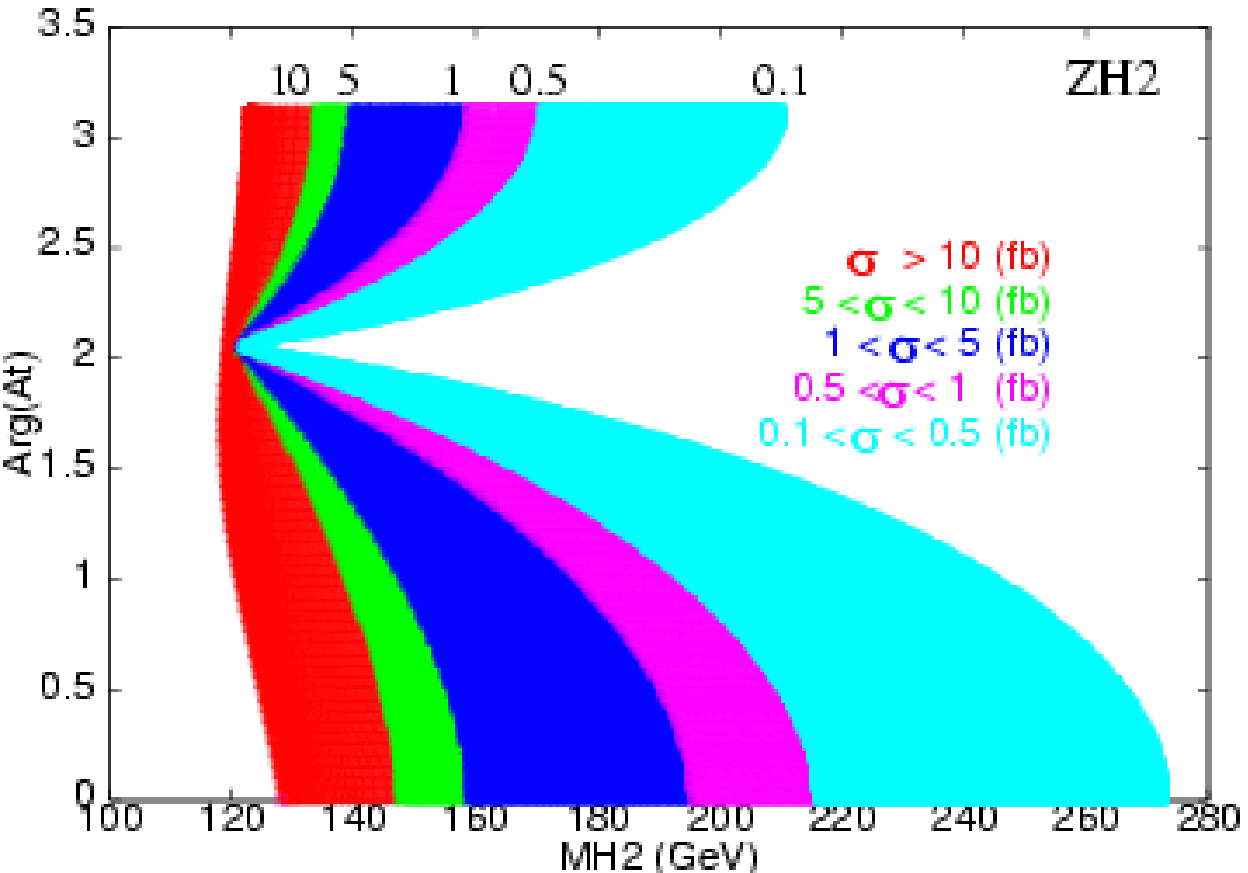}}}
\smallskip\smallskip
\centerline{{
\epsfxsize3.8 in 
\epsffile{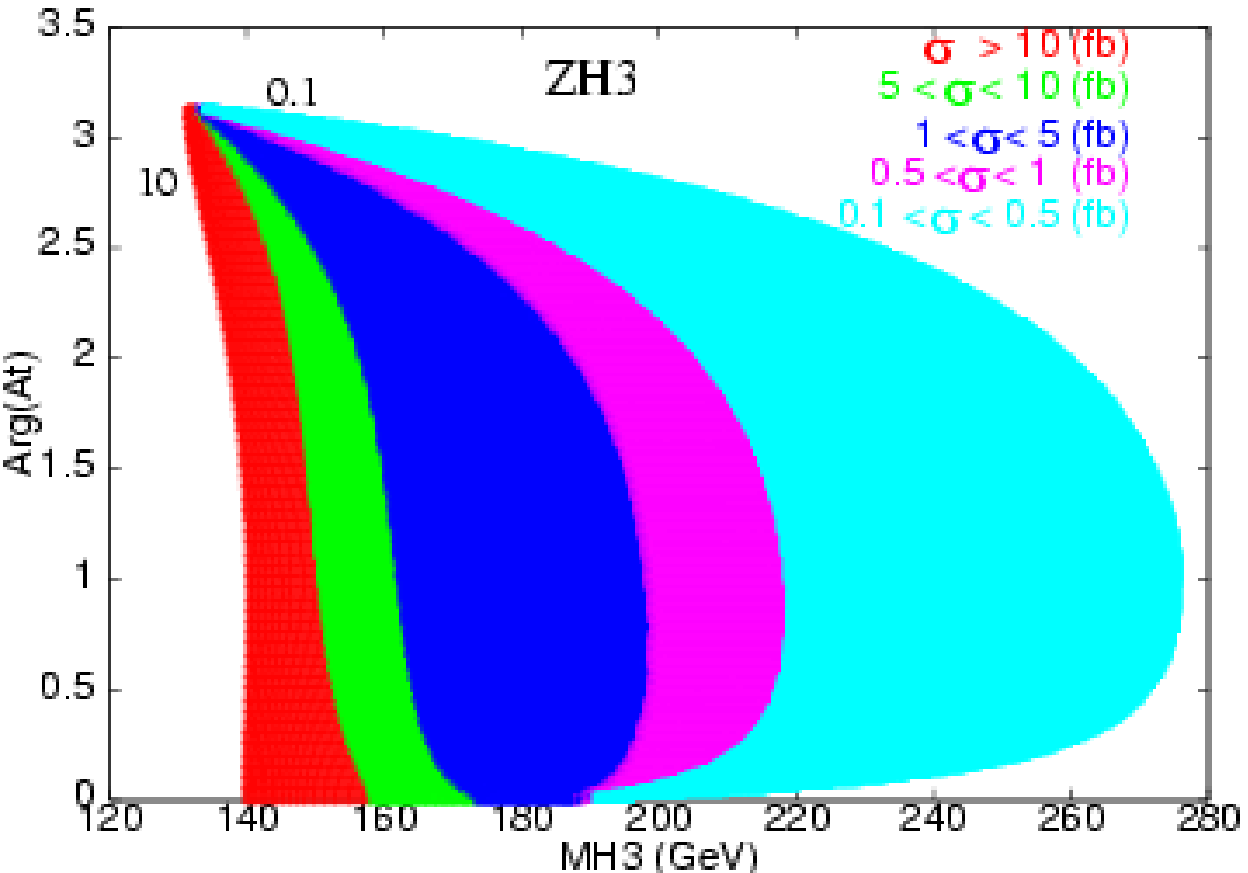}}
\hskip0.4cm
            {\epsfxsize3.8 in \epsffile{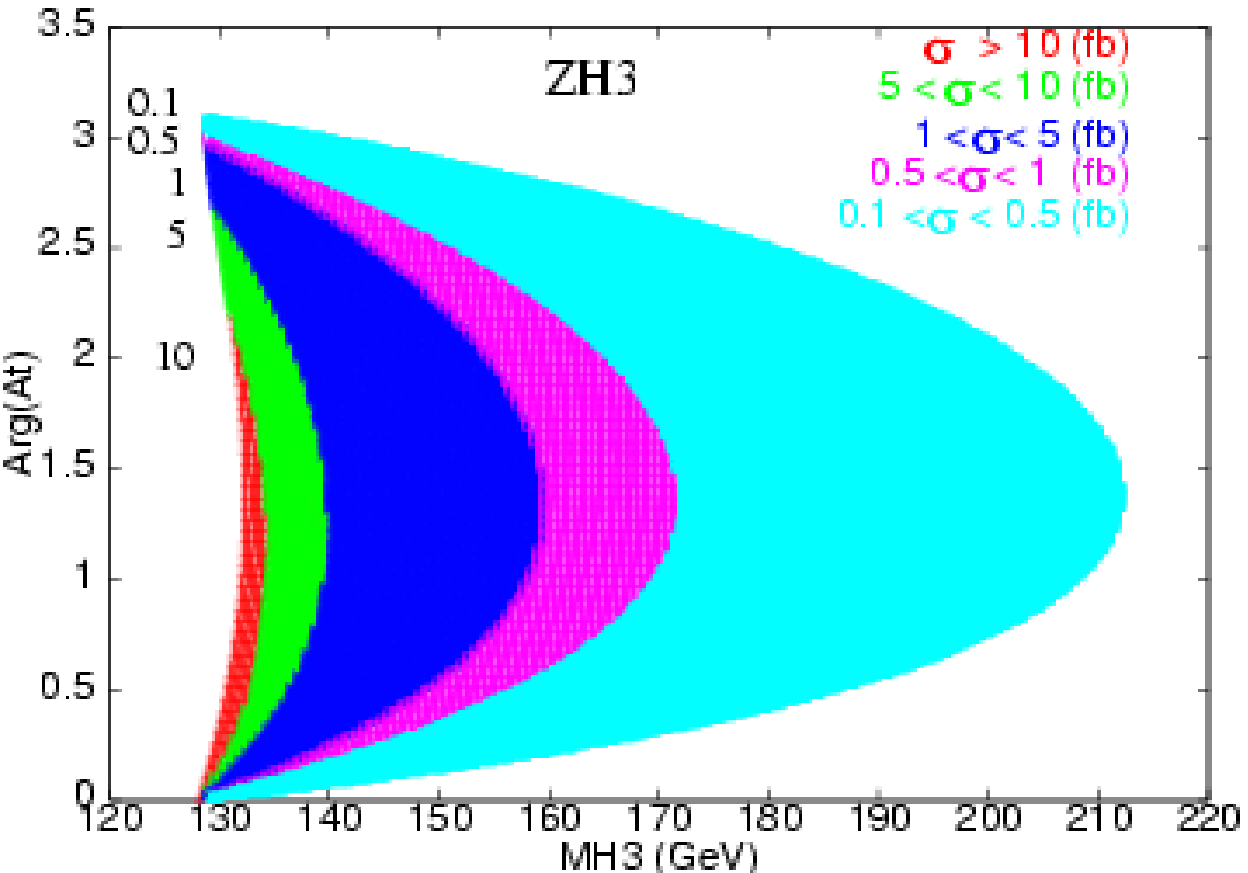}}}
\smallskip\smallskip
\caption{$\sigma(e^+e^- \to ZH^0_i)$ at $\sqrt{s}=500$ GeV 
in ($M_{H_i}$, Arg($A_t$) plane; 
$M_{SUSY}=1$ TeV, $\tan\beta=6$ (left panels) and 
$M_{SUSY}=500$ GeV, $\tan\beta=15$ (right panels)}
\label{cros2}
\end{figure}

\newpage

\begin{figure}[t!]
\centerline{{
\epsfxsize3.8 in 
\epsffile{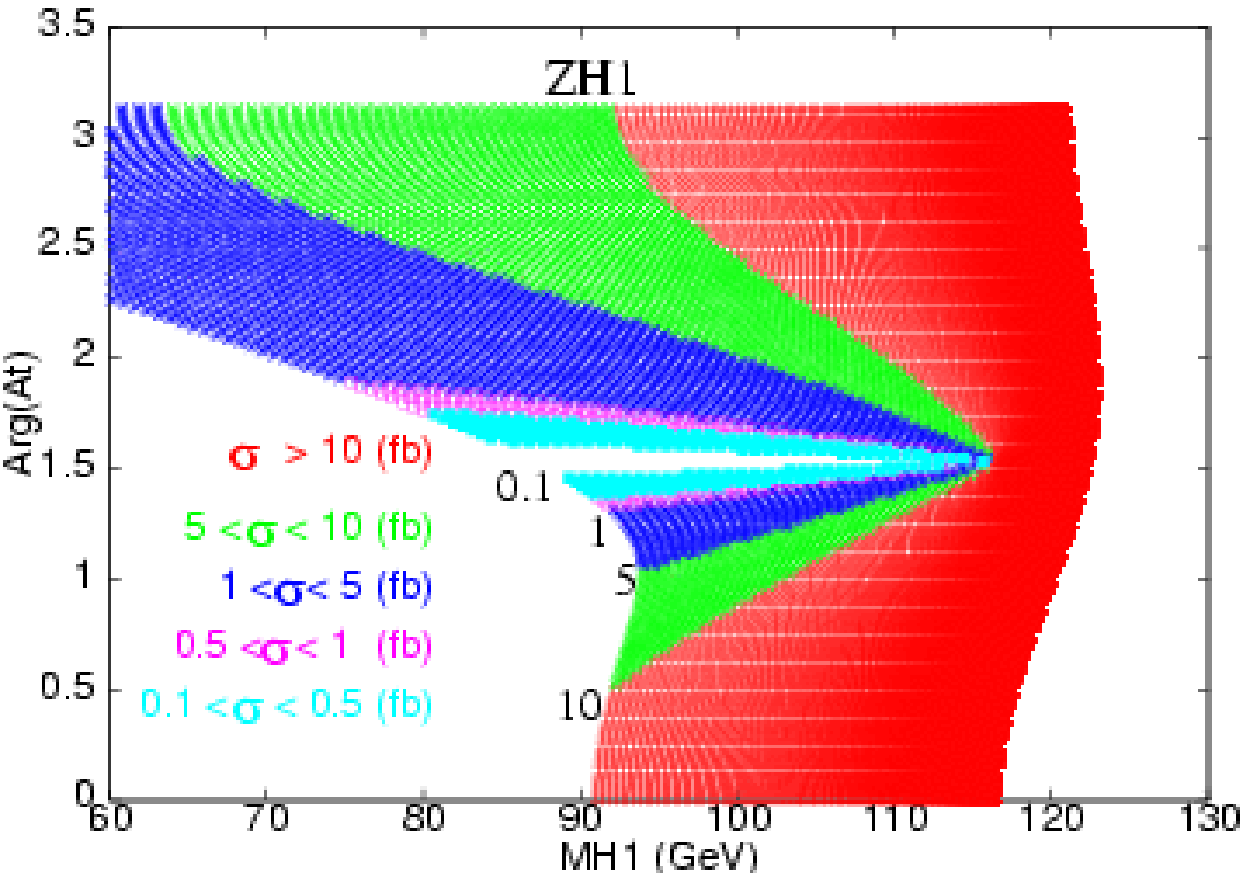}}
\hskip0.4cm
            {\epsfxsize3.8 in 
\epsffile{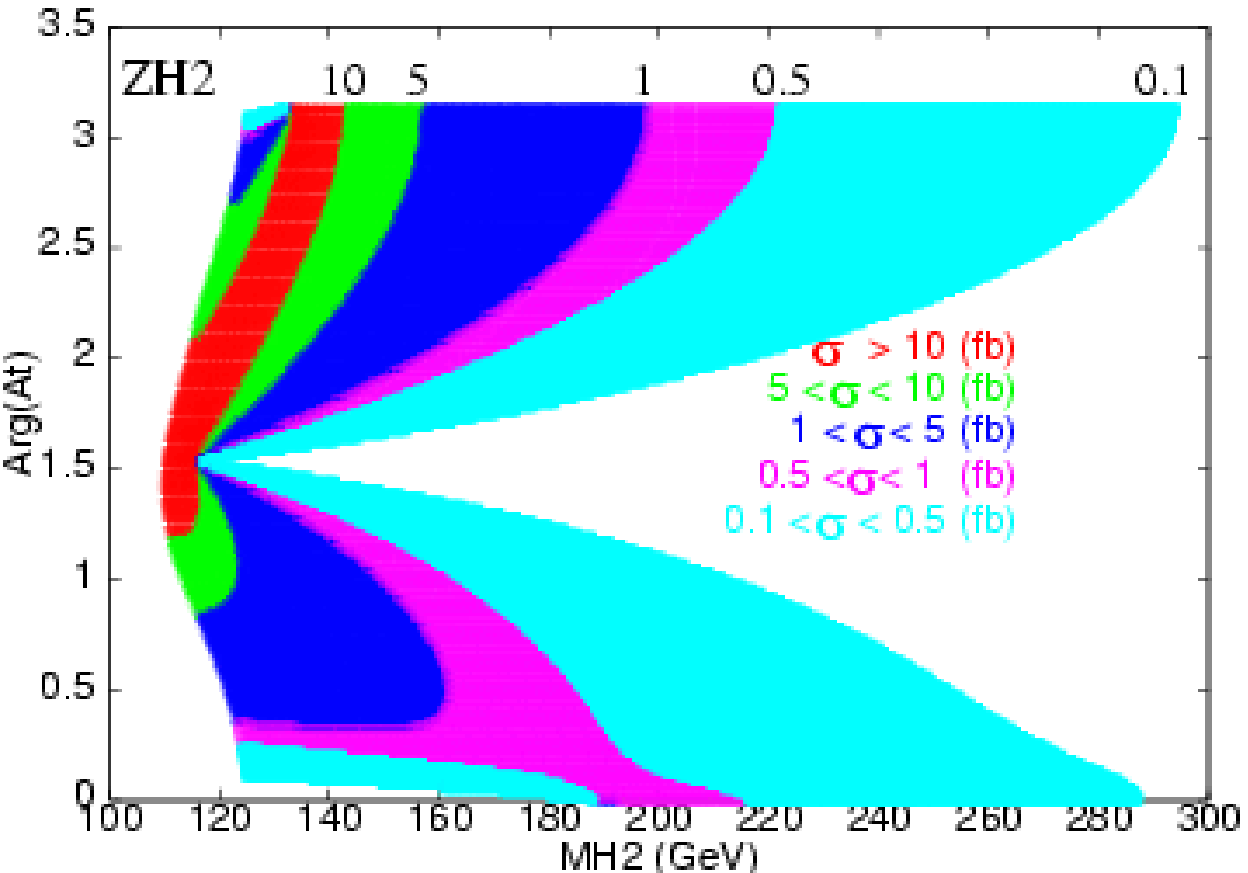}}}
\smallskip\smallskip 
\centerline{{
\epsfxsize3.8 in 
\epsffile{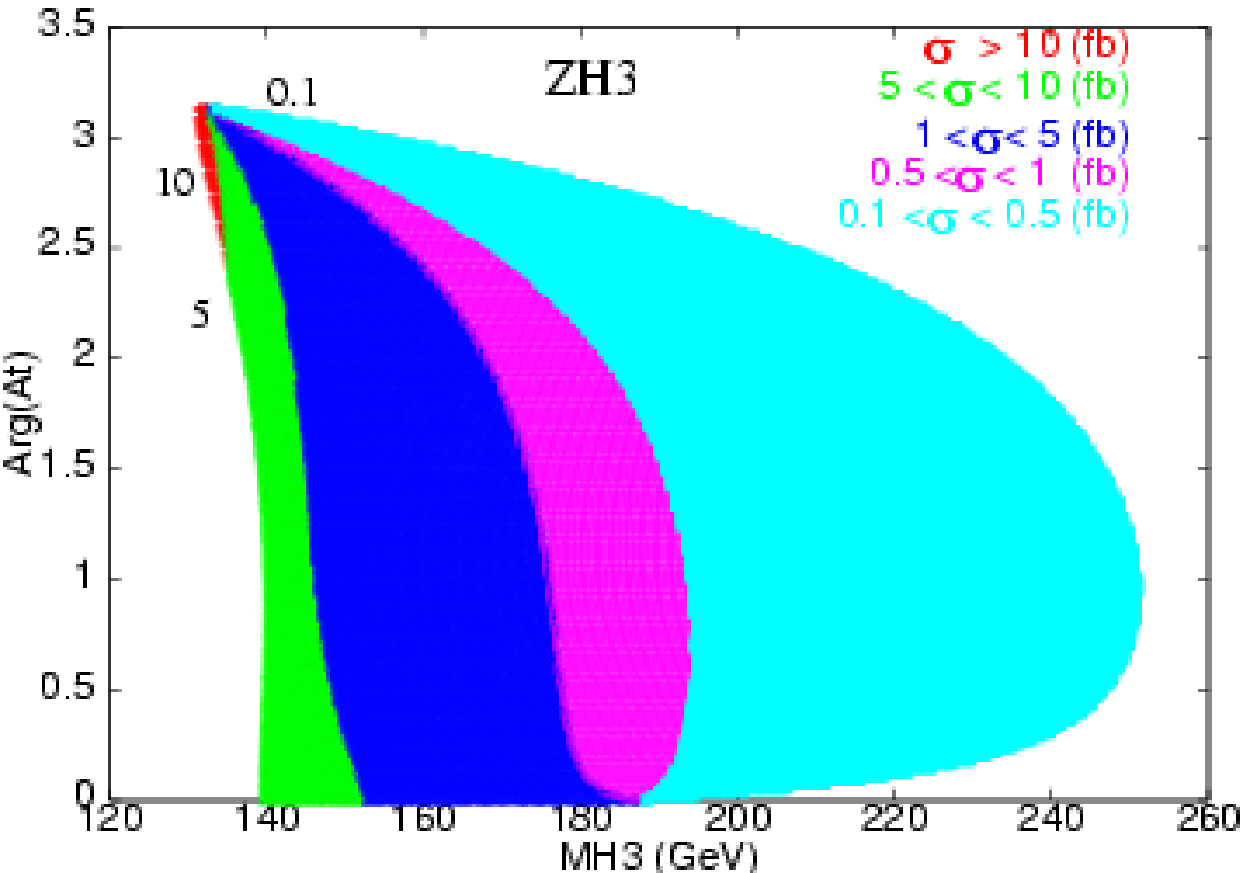}}}
\caption{$\sigma(e^+e^- \to ZH^0_i)$ at $\sqrt{s}=800$ GeV 
in ($M_{H_i}$, Arg($A_t$) plane for $M_{SUSY}=1$ TeV, $\tan\beta=6$ }
\end{figure}

\newpage
\begin{figure}[t!]
\smallskip\smallskip 
\centerline{{
\epsfxsize3.8 in 
\epsffile{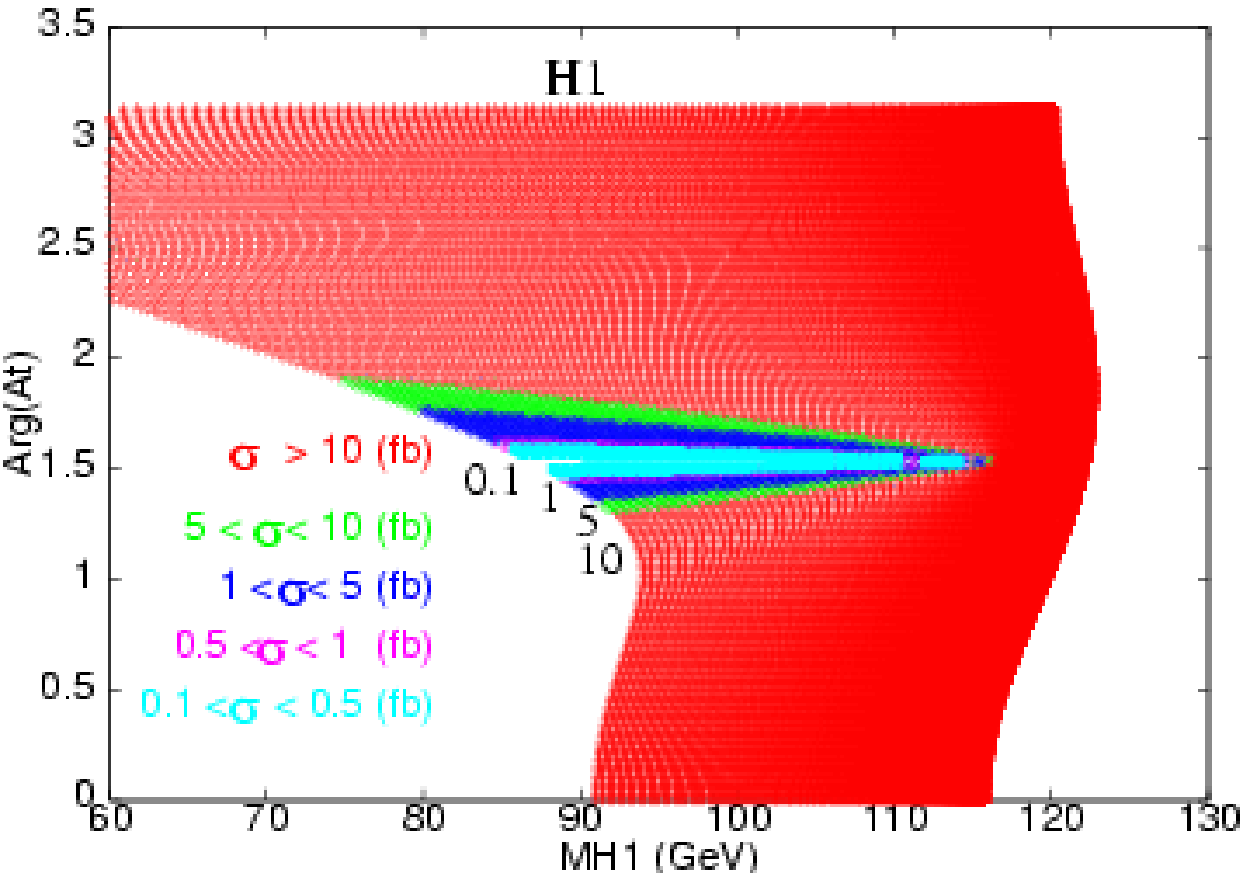}}
\hskip0.4cm
            {\epsfxsize3.8 in \epsffile{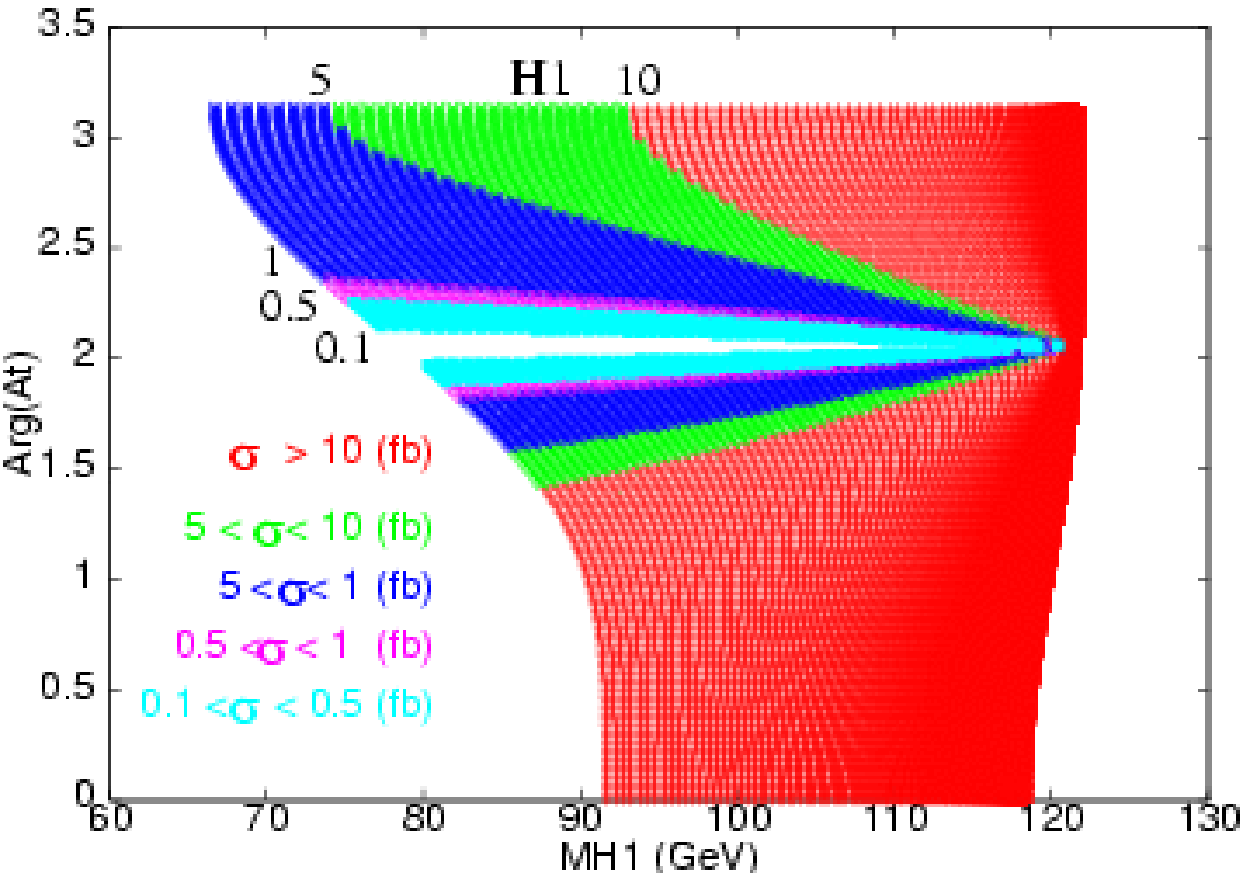}}}
\smallskip\smallskip
\centerline{{
\epsfxsize3.8 in 
\epsffile{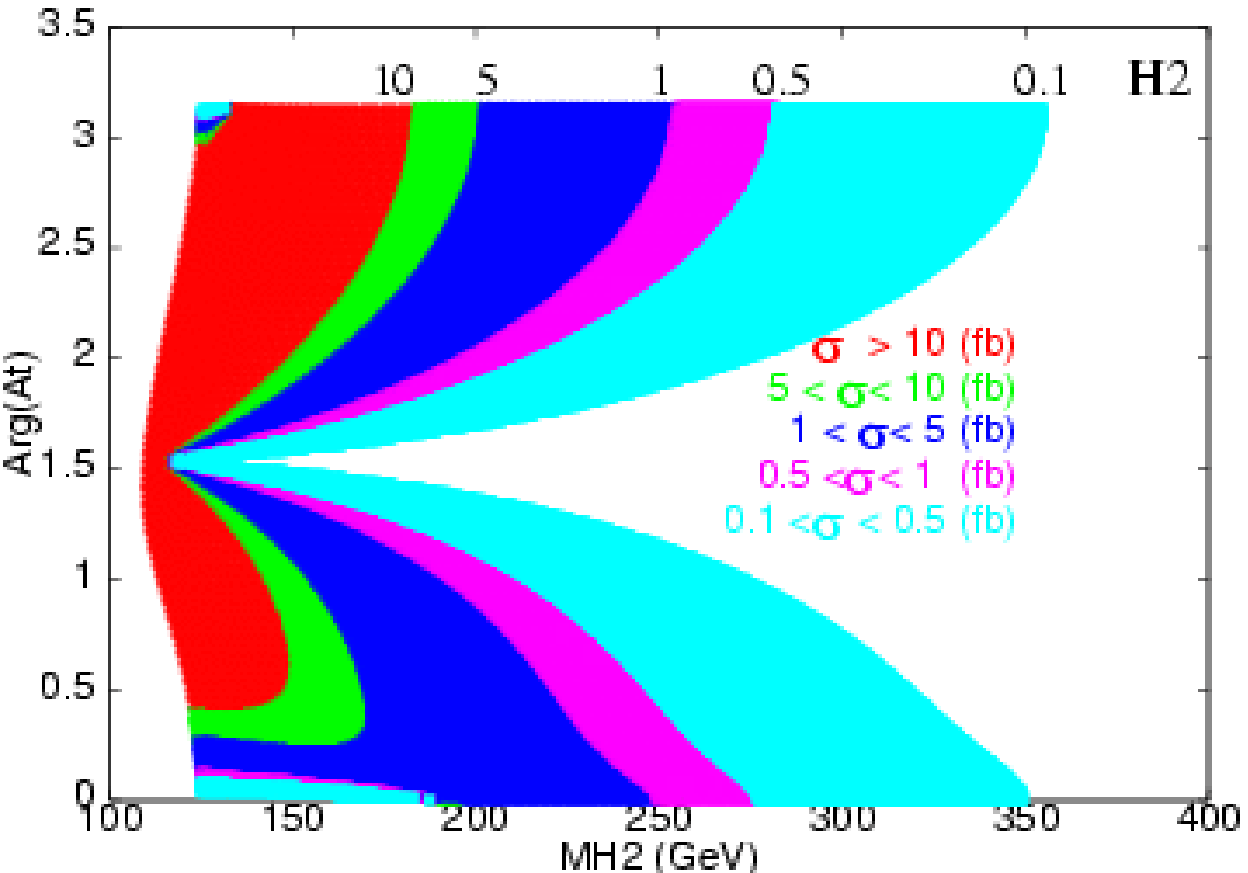}}
\hskip0.4cm
            {\epsfxsize3.8 in \epsffile{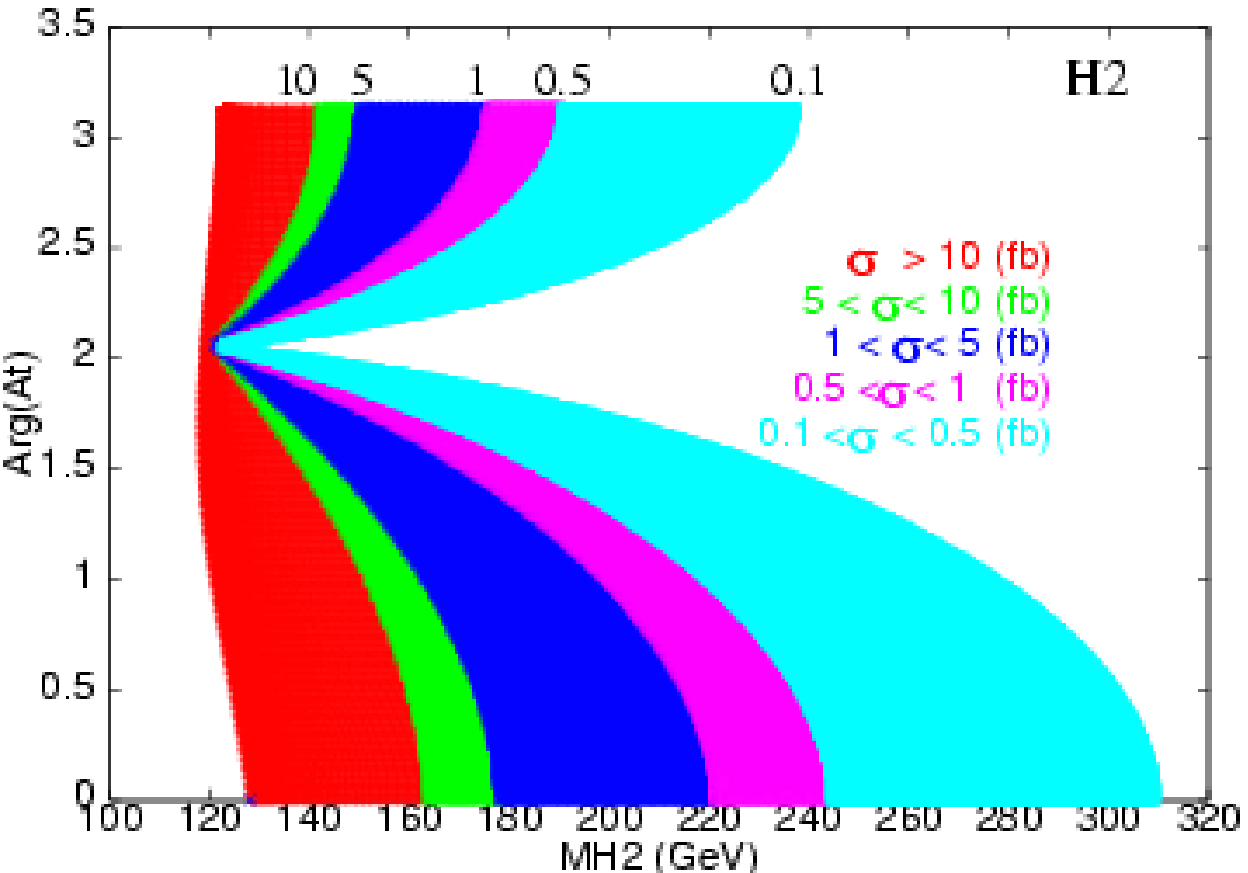}}}
\smallskip\smallskip
\centerline{{
\epsfxsize3.8 in 
\epsffile{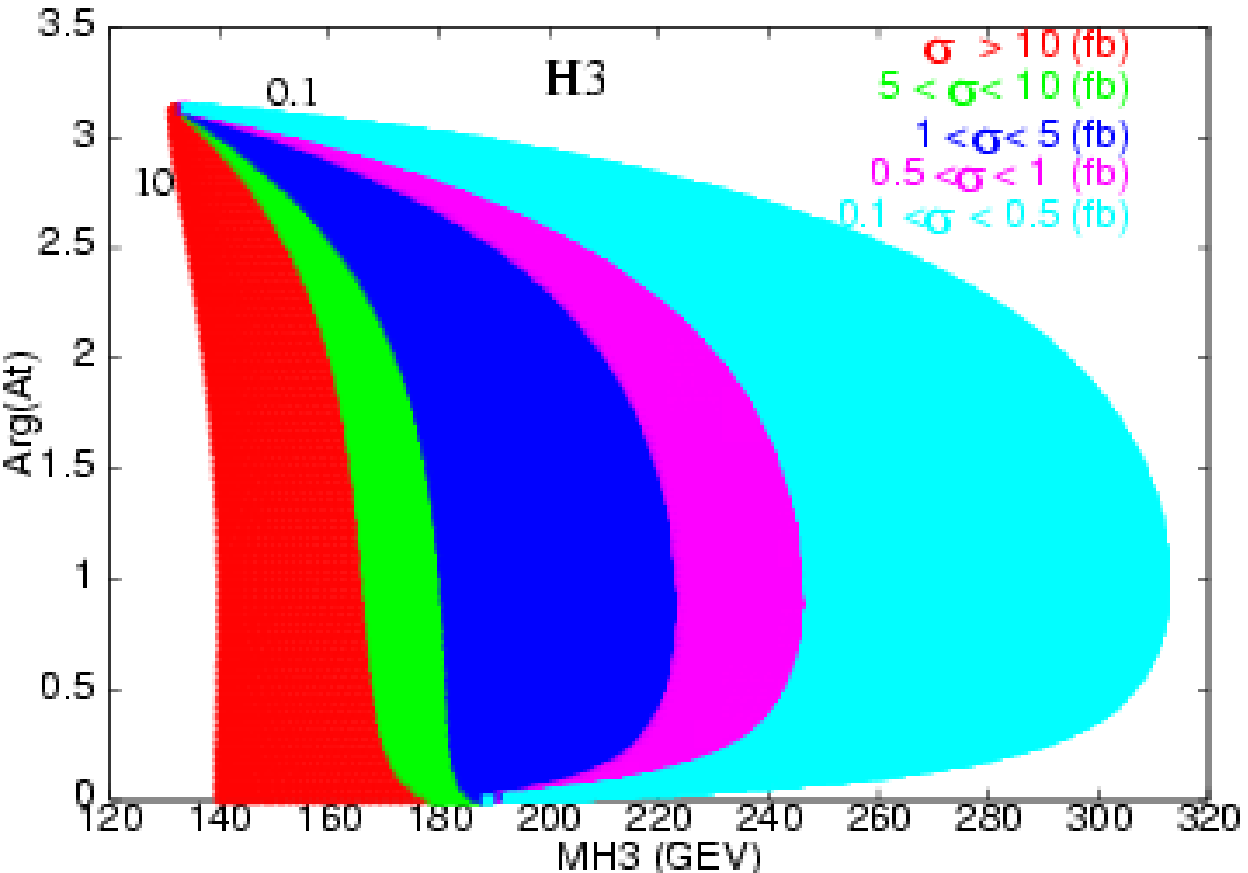}}
\hskip0.4cm
            {\epsfxsize3.8 in \epsffile{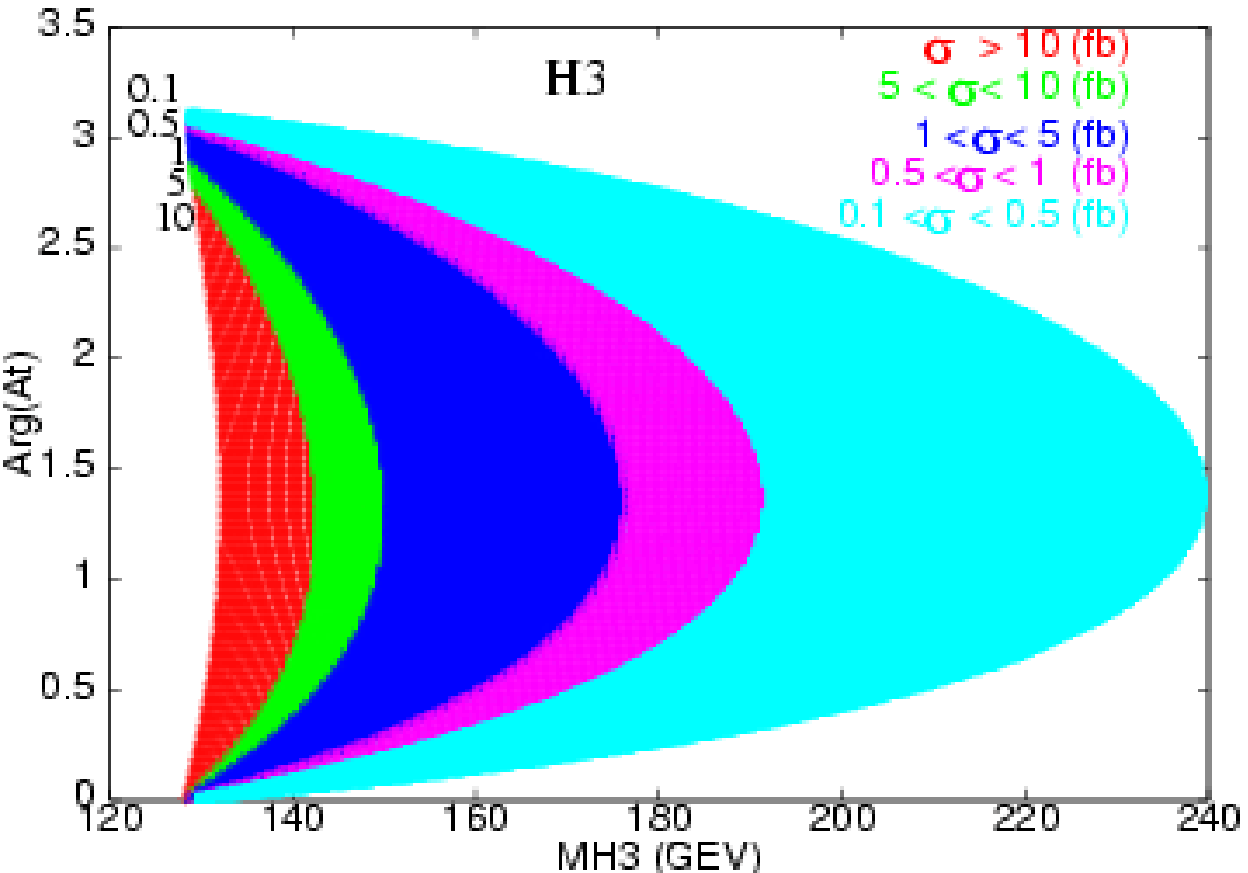}}}
\smallskip\smallskip
\caption{$\sigma(e^+e^- \to H^0_i \nu\overline{\nu})$ at 
$\sqrt{s}=800$ GeV in ($M_{H_i}$, Arg($A_t$) plane, 
$M_{SUSY}=1$ TeV; $\tan\beta=6$ (left panels) and 
$M_{SUSY}=500$ GeV, $\tan\beta=15$ (right panels)}
\label{cros2}
\end{figure}

\newpage

\begin{figure}[t!]
\smallskip\smallskip 
\centerline{{
\epsfxsize3.8 in 
\epsffile{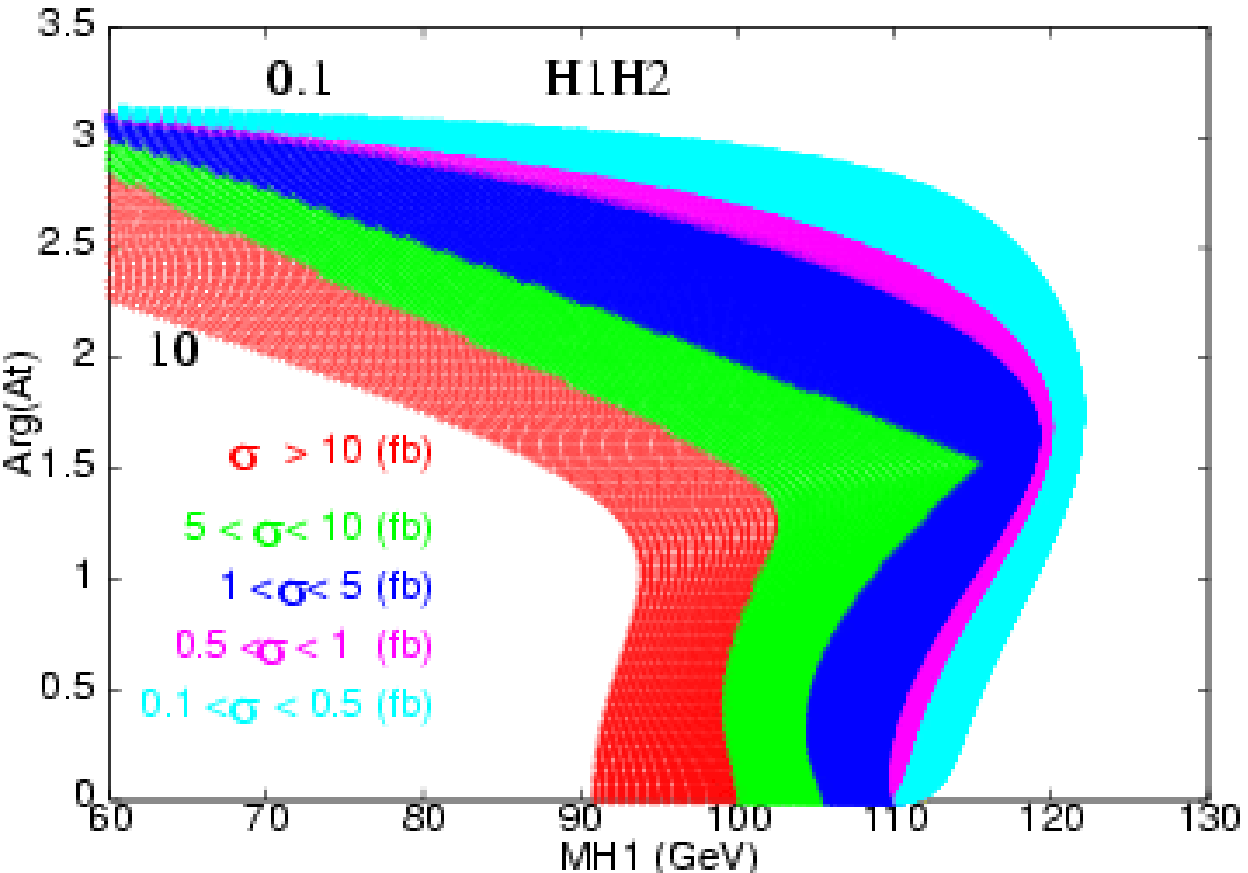}}
\hskip0.4cm
            {\epsfxsize3.8 in \epsffile{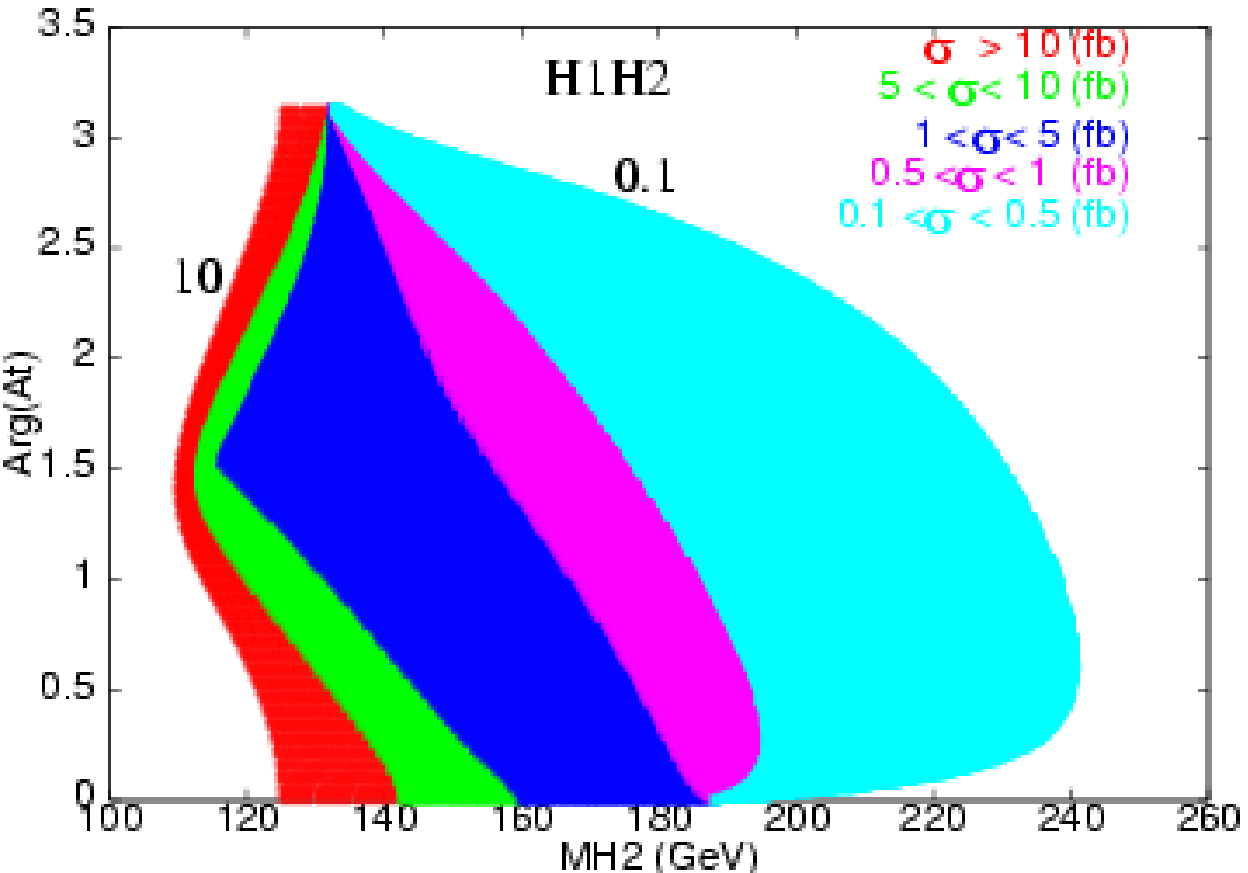}}}
\smallskip\smallskip
\centerline{{
\epsfxsize3.8 in 
\epsffile{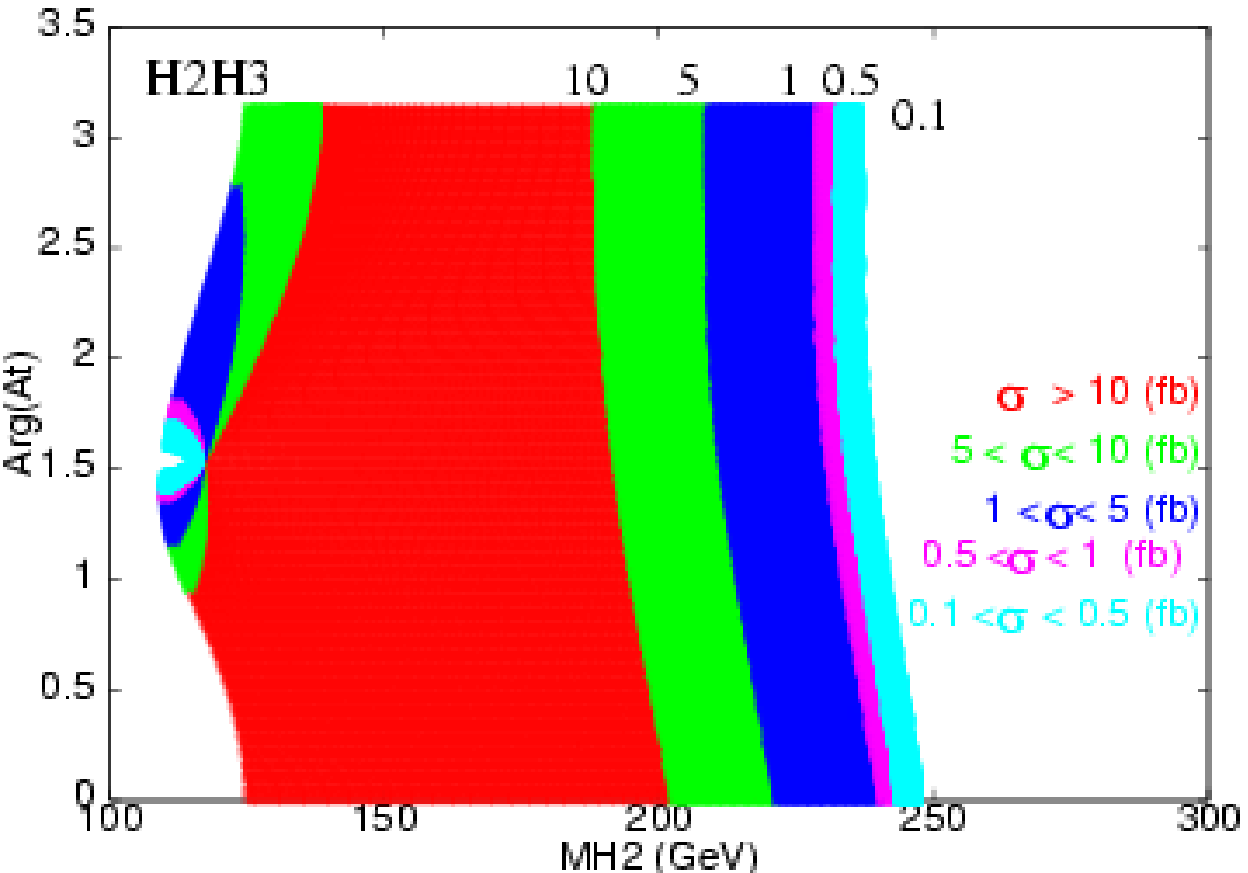}}
\hskip0.4cm
            {\epsfxsize3.8 in \epsffile{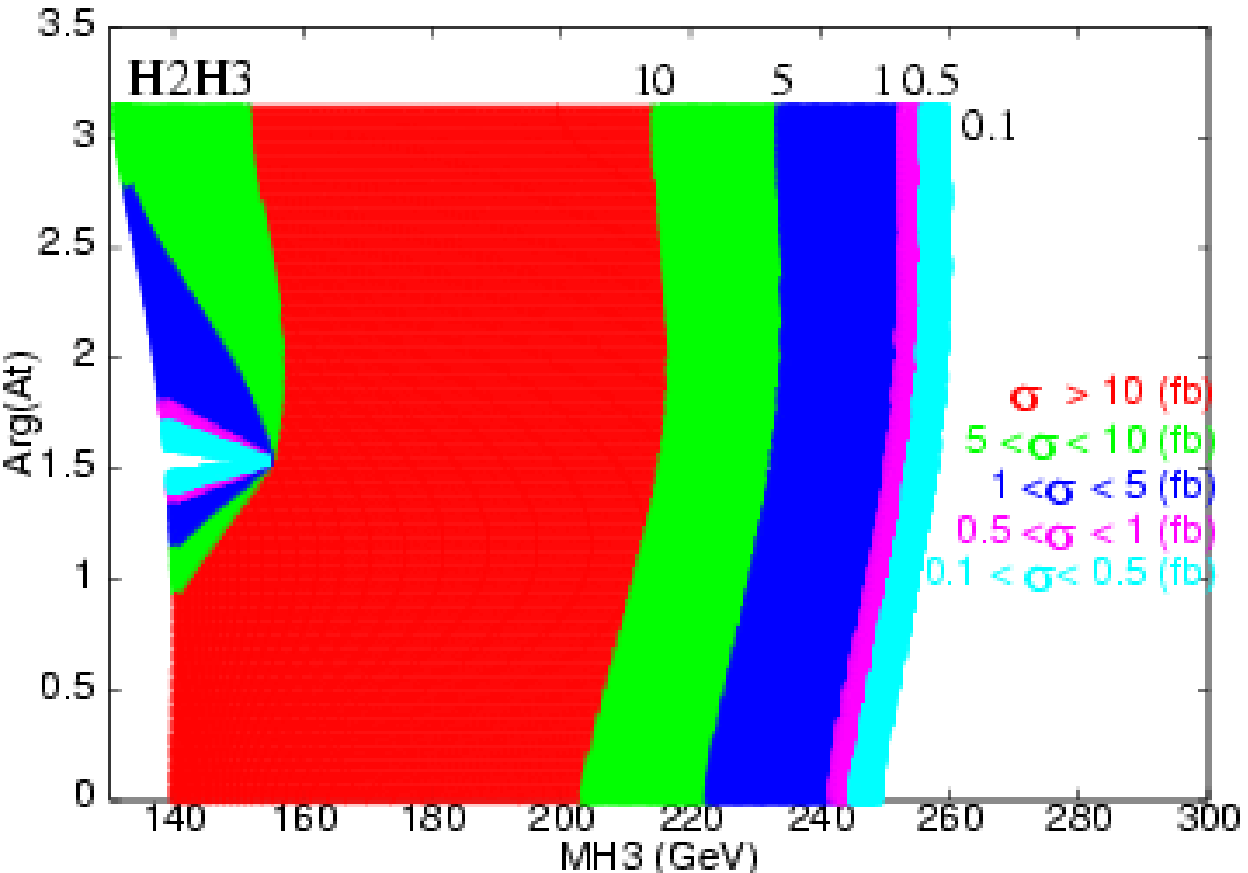}}}
\smallskip\smallskip
\centerline{{
\epsfxsize3.8 in 
\epsffile{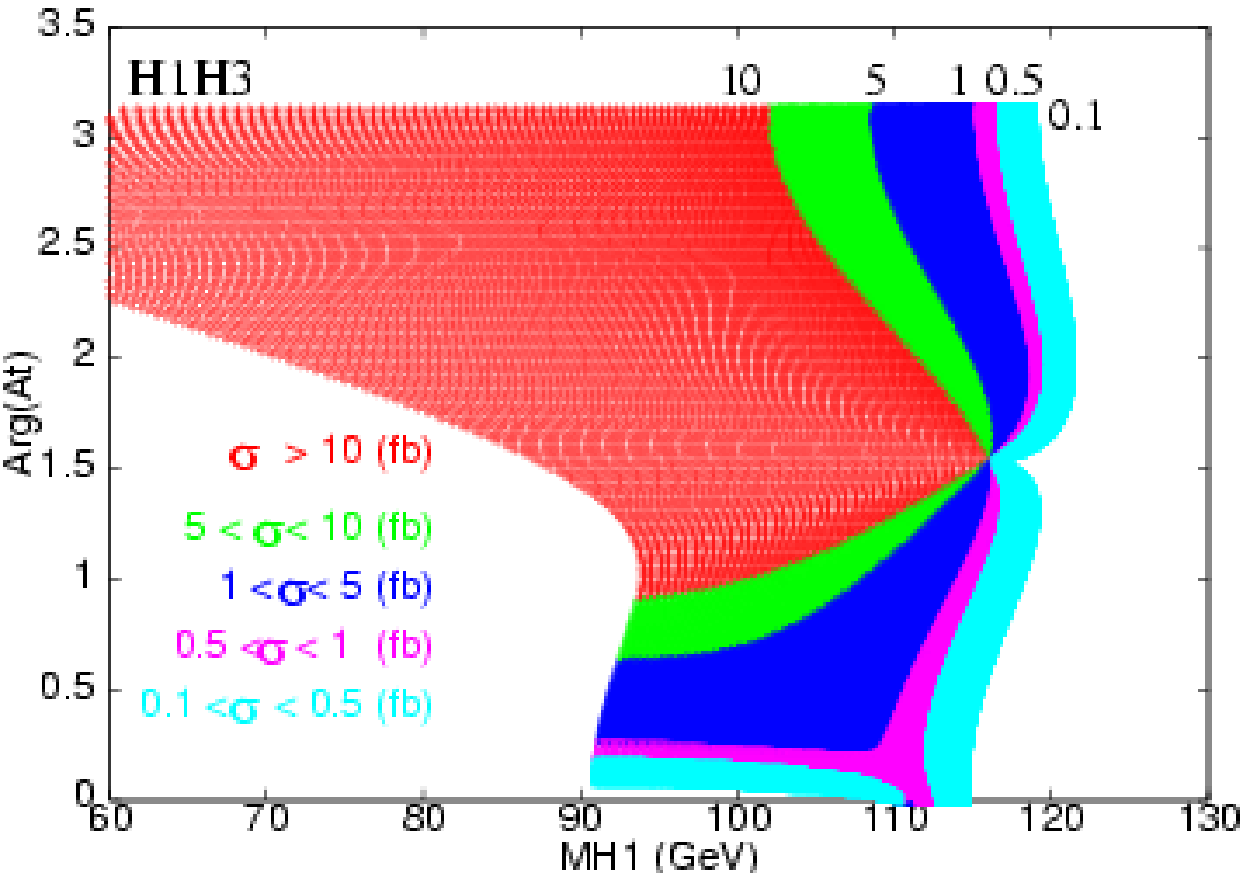}}
\hskip0.4cm
            {\epsfxsize3.8 in \epsffile{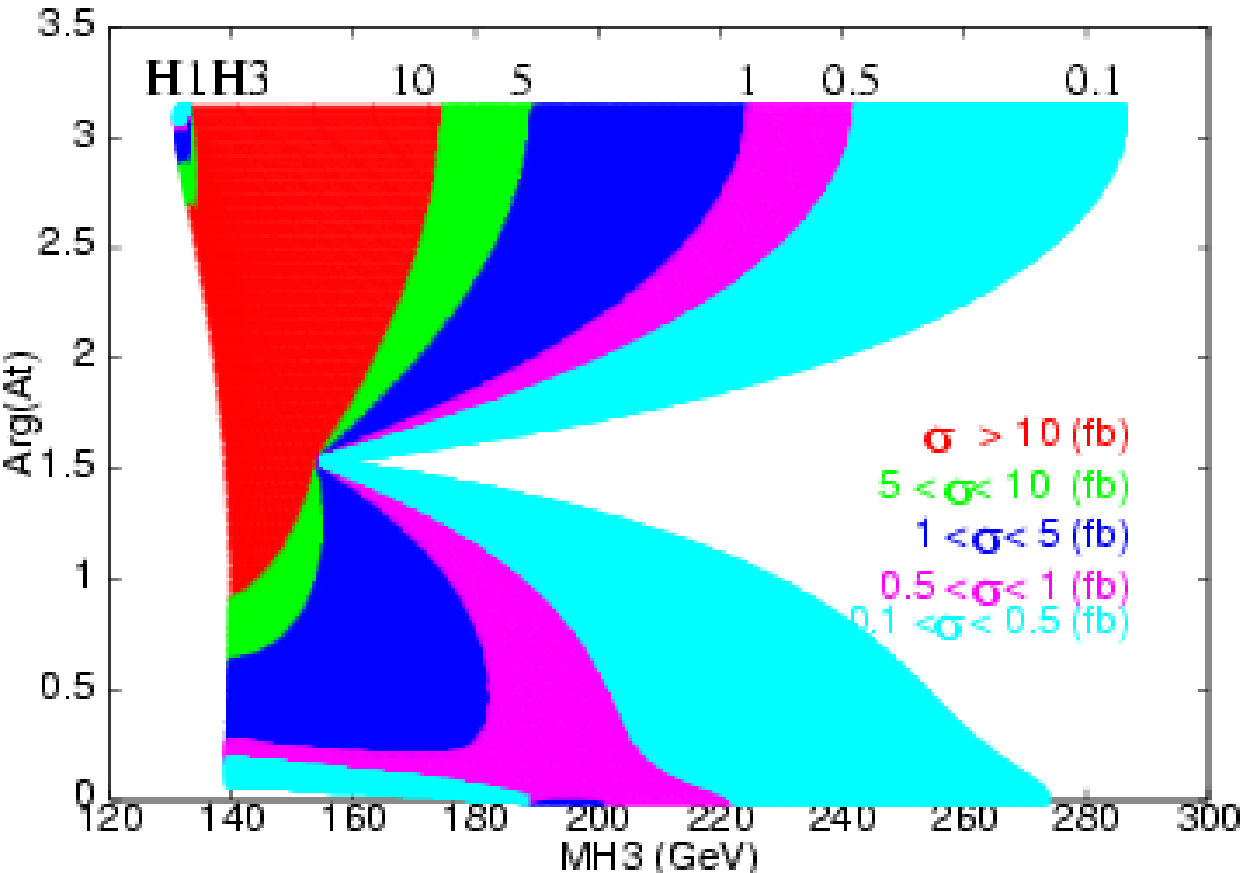}}}
\caption{$\sigma(e^+e^-\to H^0_i H^0_j)$ at $\sqrt{s}=500$
in ($M_{H_{i,j}}$, Arg($A_t$) plane, $M_{SUSY}=1$ TeV, 
$\tan\beta=6$}
\end{figure}

\end{document}